# Predicting densities and elastic moduli of $SiO_2$-based glasses by machine learning

Yong-Jie Hu[1], Ge Zhao[2], Mingfei Zhang[1], Bin Bin[1], Tyler Del Rose[1], Qian Zhao[3], Qun Zu[3], Yang Chen[3], Xuekun Sun[4], Maarten de Jong[5,6] and Liang Qi[1*]

[1]Department of Materials Science and Engineering, University of Michigan, Ann Arbor, Michigan, 48109, USA

[2]Department of Statistics, Pennsylvania State University, State College, Pennsylvania,16802, USA

[3]Sinoma Science & Technology Co., Ltd., Nanjing, Jiangsu, 210012, China

[4]Continental Technology LLC, Indianapolis, Indiana, 46033, USA

[5]Department of Materials Science and Engineering, University of California, Berkeley, California, 94720, USA

[6]Space Exploration Technologies (SpaceX), Hawthorne, California, 90250, USA

*corresponding author: qiliang@umich.edu

**Abstract**

Chemical design of $SiO_2$-based glasses with high elastic moduli and low weight is of great interest. However, it is difficult to find a universal expression to predict the elastic moduli according to the glass composition before synthesis since the elastic moduli are a complex function of interatomic bonds and their ordering at different length scales. Here we show that the densities and elastic moduli of $SiO_2$-based glasses can be efficiently predicted by machine learning (ML) techniques across a complex compositional space with multiple (>10) types of additive oxides besides $SiO_2$. Our machine learning approach relies on a training set generated by high-throughput molecular dynamic (MD) simulations, a set of elaborately constructed descriptors that bridges the empirical statistical modeling with the fundamental physics of interatomic bonding, and a statistical learning/predicting model developed by implementing least absolute shrinkage and selection operator with a gradient boost machine (GBM-LASSO). The predictions of the ML model are comprehensively compared and validated with a large amount of both simulation and experimental data. By just training with a dataset only composed of binary and ternary glass samples, our model shows very promising capabilities to predict the density and elastic moduli for k-nary $SiO_2$-based glasses beyond the training set. As an example of its potential applications, our GBM-LASSO model was used to perform a rapid and low-cost screening of many (~$10^5$) compositions of a multicomponent glass system to construct a compositional-property database that allows for a fruitful overview on the glass density and elastic properties.

# 1. Introduction

$SiO_2$-based glasses are a group of materials known for its diverse applications as both structural and functional materials in various industrial fields[1-3]. Density and elastic moduli are two of the most common properties of $SiO_2$-based glasses. Particularly, discovering new glass compositions to achieve high elastic moduli and low densities is of great interests for the development of strengthened and durable $SiO_2$-based glass materials nowadays. Finding universal expressions or correlations to efficiently predict and further optimize densities and elastic moduli of $SiO_2$-based glasses according to the chemical composition is not very straightforward due to their non-crystalline structures. Different from the crystalline materials, the elastic moduli of a $SiO_2$-based glass are not only determined by the atomic bonding strength but also a complex function of many other physical properties at different length scales[4-7], such as cation coordination, formation of atomic ring, chain, layer and polyhedral atomic clusters, and even the structural organization at mesoscopic scale, e.g. the formation of nanodomains[4]. Moreover, the additive oxides besides $SiO_2$ introduce cations with various valence states, which not only change the cation-oxygen bonding strengths but also modify the degree of network polymerization. As a result, elastic moduli of the glass are complex functions of the chemical compositions of the additive oxides.

Through linear or polynomial regression analyses, many efforts have been devoted previously to fit the densities and elastic moduli with either the glass composition only[8–10] or a single parameter related to atomistic structures, such as molar volume[11] and the correlation length of x-ray diffraction peak[5,12]. Although these regression models were demonstrated to provide valid descriptions for some certain glass systems, they may have two major shortcomings that impede their usage in the practical design of new glass compositions. Firstly, the models are usually accurate for specific glass systems. Once the type of the additive oxides changed, the regression results may significantly be varied, or an alternative modeling method must be applied. As a result, it is difficult to extrapolate the developed models to capture the mixed effects of multiple additive oxides in the design space for industrial glass products. Secondly, for the models built on non-compositional variables, their outcomes are hard to be directly used for discovering new glass compositions, because it could be difficult to quantitatively interpret the optimization results with respect to glass chemistries. For example, elastic extremeness may occur at a certain correlation length of x-ray diffraction peak[5,12], while it is still unknown what glass chemistries result in such correlation length. These shortcomings may originate from the fact that these models were usually built from regression algorithms based on presumed analytical formulas and a few variables that were predetermined relying on historical intuition and knowledge.

Machine learning (ML) techniques offer an alternative way to create predictive models that bridge the materials property of interest with its potential descriptors quickly and automatically[13–16]. In addition, the model created from ML does not require to rely on presumed fitting expressions or any historical intuition of material behaviors. As a result, the ML approaches can be a particularly powerful tool for modeling the property that is determined by many factors in a complex way with unclear underlying mechanisms. To date, the ML approaches have been widely used to build predictive models for a handful of materials properties and applications, including the modeling of elastic moduli of both crystalline[17–19] and amorphous materials[20–24]. Using the artificial neural networks and genetic evolution algorithms, Mauro et al.[20,24] recently showed that Young's moduli of over 250 different glass samples can be accurately regressed and predicted using glass compositions as inputs. Most recently, by using glass composition as input descriptors, Yang et al. performed extensive studies to show that Young's moduli of the $CaO-Al_2O_3-SiO_2$ ternary glass system can be accurately predicted through several different ML models[21]. Additionally, in a recent work by Bishnoi et al., Young's moduli of four important ternary glass systems were comprehensively studied and well predicted based on nonparametric ML regression models[22]. All these recent works show great promise in the application of ML techniques on the chemistry design of advanced glass materials.

One could encounter several challenges to model densities and elastic moduli of $SiO_2$-based glasses under a ML-based framework. A typical one would be the availability of sufficient quantities of training data to sample the predictive space. It could be harmful for extrapolative predictions if the training data are clustered around one or several particular regions of the design space. However, unfortunately, experimental data are usually clustered due to the constraints of practical manufacturing. This situation can be overcome by employing atomistic simulations such as molecular dynamics (MD) and molecular statics (MS) simulations, which were proved to be able to accurately compute the elastic moduli of many glassy systems[5,7,25]. Particularly, the MD simulations offer a promise of being able to predict the elastic moduli for the glass compositions that have not been experimentally synthesized[20,26]. As a result, one can achieve a compositionally homogenous sampling for any glass system of interest without the need of concerning the practical manufacturing constraints. However, even though the MD simulation is an effective and efficient tool, with current and near-term computing techniques, it can only access a limited fraction of discrete compositions in a practical design space that contains several (~5) oxide-components using tens of millions of CPU hours. Therefore, from the practical view, it is expected that the developed ML model is capable of giving reliable predictions over a large and even the entire compositional space despite the training is based on a limit set of data of lower-order systems (e.g., binary and ternary $SiO_2$-based systems). To achieve this goal, the model cannot be purely empirical. A subtle set of descriptors should be constructed to include not only the information of glass composition but also the physical information related to the chemical characteristic of the components[26], such as the parameters associated with atomic bond energies. In fact, several recently developed physic-based topological models have demonstrated quantitative connections between glass elasticity and the free energies associated with breaking different bond constraints between cations and anions[27,28].

In this work, through merging ML approaches with high-throughput MD simulations, we aimed to develop a quantitatively accurate model to predict densities and elastic moduli of $SiO_2$-based glasses according to the glass composition but across a complex compositional space. The effects of 11 types of additive oxides were investigated, namely $Li_2O$, $Na_2O$, $K_2O$, CaO, SrO, $Al_2O_3$, $Y_2O_3$, $La_2O_3$, $Ce_2O_3$, $Eu_2O_3$ and $Er_2O_3$. The training set was generated using MD simulations to homogenously sample the density and elastic properties of a part of the constituent binary and ternary systems. A set of descriptors was carefully constructed from the force-field potentials used for MD simulations and elemental mole fractions to include both physical and compositional information. Sequentially, enlightened by the previous work[17], a statistical learning/predicting model was developed by implementing the least absolute shrinkage and selection operator[29] with a gradient boost machine[30] (GBM-LASSO). As a comparison, a traditional decision tree-based model[31,32] was also employed. By validating with a large amount (>>1000) of both simulation and experimental data, the GBM-LASSO model was demonstrated to have promising prediction capabilities on both densities and elastic moduli for the $SiO_2$-based glasses not only within the composition range of the training set but also the high-dimension compositional spaces beyond the training set. The developed ML model could be useful for rapid glass composition-property screening that allows for a fruitful estimation and overview on the density and elastic properties of the general multi-component glass systems, especially the novel composition regions.

## 2. Details of MD simulations

To establish the training set, high-throughput MD simulations followed by energy minimizations were employed to calculate the density, bulk and shear moduli over 498 different glass compositions. The compositions were from 11 binary and 20 ternaries systems, which are specified in Table 1. For each system, the mole fractions of the additive oxides species were varied from 0 mol% to 35 mol% for every 5 mol%, while the composition of $SiO_2$ in the systems was kept no less than 65 mol%. For example, for binary systems, calculations were performed at seven compositions, which are $xA_nO_m$-(100-$x$)$SiO_2$ with x =0, 5, 10, 15, 20, 25, 30 and 35 mol%, respectively. For ternary systems, in addition to the compositions already calculated in constituent binary systems, calculations were performed at the compositions of $xA_nO_m$-$yB_kO_l$-(100-$x$-$y$)$SiO_2$, where x,y=5, 10, 15, 20, 25, and 30 mol% and x+y ≤ 35 mol%. $A_nO_m$ and $B_kO_l$ represent the additive oxides species.

In the present work, the MD simulations were performed using a set of interatomic potentials developed by Du and Cormack[25,33–42], which are found to yield reliable predictions on the densities and elastic moduli of various $SiO_2$-based glasses[25,43–46]. Another advantage of this potential set is that it covers the common oxides that include most of the industrial glass components. The potential consists of long-range Coulombic interactions and short-range interactions described in the Buckingham form[47], which can be expressed as,

Equation 1
$$U_{i,j}(r_{i,j}) = \frac{q_i q_j}{4\pi\varepsilon_0 r_{i,j}} + A_{i,j}\exp\left(-\frac{r_{i,j}}{B_{i,j}}\right) - \frac{C_{i,j}}{r_{i,j}^6}$$

where $r_{i,j}$ is the interatomic distance between atom $i$ and $j$, $q_i$ and $q_j$ are the effective ionic charges of atom $i$ and $j$, respectively, and $A_{i,j}$, $B_{i,j}$ and $C_{i,j}$ are the energy parameters of the Buckingham form between $i$ and $j$. In this set of potential, the short-range interactions between cations are not considered since it is assumed that two cations cannot be the first-nearest neighbor ions/atoms. The values of the effective ionic charges and Buckingham parameters for each element are summarized in Table 2. Moreover, it should be noted that, by following the method developed by Deng and Du[45], one of the Buckingham parameters of the boron (B) ion, $A_{B,O}$, was varied with the glass composition in each MD simulation in order to capture the changes in the partitioning between the $BO_3$ and $BO_4$ clusters caused by different chemical environments.

All the MD simulations were performed using the LAMMPS package[48]. Coulomb interactions were evaluated by the Ewald summation method, with a cutoff of 12 Å. The cutoff distance of the short-range interactions was chosen to be 8.0 Å. Cubic simulation boxes were constructed to consist of about 2100 atoms so that the mole fraction of each oxide species of the samples in the training set can be achieved. Initial atomic coordinates were randomly generated using the program PACKMOL[49]. The simulation protocol was initiated with relatively equilibration runs of 0.5 ns at 5000 K to remove the memory effects of the initial structure, followed by a linear cooling procedure with a nominal cooling rate of 10K/ps to 3000 K in the canonical (NVT) ensemble. Then, the system was further equilibrated for 0.5 ns at 3000K in the isothermal–isobaric ensemble (NPT with zero pressure) to allow a relaxation of the simulation box and atomic positions simultaneously. After this, a MD run with the microcanonical ensemble (NVE) was performed for another 0.5 ns to further equilibrate the system. After the equilibration at 3000K, the system was gradually cooled down to 300 K through steps of 2500 K, 2000 K, 1000 K, 300 K with a nominal cooling rate of 0.5 K/ps under NPT condition. At each step temperature, the system was equilibrated for 0.5 ns under NPT condition, and then run with an NVE ensemble for another 0.5 ns. At 300 K, the system is equilibrated for 1 ns under NPT condition, which is then followed by a 0.5 ns NVE run. During the final 500,000 NVE steps, atomic configurations were recorded every 1000 steps, and an average of the configurations was taken every 50 records. Eventually, 10 (10 = 500,000/1000/50) atomic configurations of each glass composition were obtained and used for the further calculations of densities and elastic moduli. Recording multiple atomic configurations would allow us to avoid accidentally using a single *unreasonable* configuration that can lead to large errors in the following energy minimization calculations.

The elastic constants $c_{ij}$ for a system are defined as the second derivative of the potential energy $U$ at the corresponding local minimum (the curvature of the potential energy) with respect to small strain deformations, $\varepsilon_i$,

Equation 2
$$c_{ij} = \frac{1}{V}\left(\frac{\partial^2 U}{\partial \varepsilon_i \partial \varepsilon_j}\right)$$

Based on the Voigt approximation[50], which provides the upper bound of elastic properties in terms of uniform strains, the bulk modulus ($K$) of the system is calculated as,

Equation 3
$$K = \frac{1}{9}(c_{11} + c_{22} + c_{33} + 2(c_{12} + c_{13} + c_{23}))$$

and the shear modulus ($G$) is calculated as,

Equation 4
$$G = \frac{1}{15}(c_{11} + c_{22} + c_{33} + 3(c_{44} + c_{55} + c_{66}) - c_{12} - c_{13} - c_{23})$$

Based on $K$ and $G$, the Young's modulus ($E$) is given by,

Equation 5
$$E = 9KG/(3K + G)$$

With the glassy structures collected from the MD simulations, the density and elastic moduli were computed by means of the GULP code[51]. The cutoffs for the Coulomb and short-range interactions were set to be same as the MD simulations. A Newton-Raphson energy minimization was performed at zero pressure and temperature to fully relax the output glassy structures from LAMMPS simulations. Then, the density was calculated theoretically by dividing the total system mass by the volume of the relaxed structure. For each glass composition, the GULP calculations were performed for all the 10 atomic structures obtained from the MD simulations, and then the average values of the density and elasticity calculations were taken as the final results. Most of the calculated elastic moduli and densities are well compared with available experimental data[1]. The results are summarized in Figures S1-S3 in Supplementary Material.

## 3. Statistical models for machine learning

### 3.1. Construction of descriptors

The successful application of ML approaches on the modeling of material properties requires the selection of an appropriate set of modeling variables or, namely, the descriptors for the property of interest. In general, the descriptors are expected to be capable of both sufficiently distinguishing each of the modeled compounds/materials and determining the targeted property. In this context, chemical compositions are straightforwardly used as one type of the most common descriptors as they are usually unique for each modeled material, and many material properties are eventually compositional dependent. In fact, several recent works have shown that using chemical compositions only as descriptors can describe the glass properties through the artificial neural network based ML algorithm[20–22,52]. However, only using compositional descriptors could make the model have limited extrapolative ability[13,24,26].

Alternatively, one can construct the descriptors using a group of material feature parameters that have physical correlations with the targeted property. In this way, the resulting model could potentially capture the underlying physical mechanisms after training, and thus offer reliable predictions for the chemistries beyond the training set. These material quantities are generally classified into two categories, namely the *chemical* and *structural* feature parameters[15,17]. *Chemical* feature parameters are usually elemental properties, such as the effective ionic charge, atomic radius and weight, and electronegativity, which can be obtained by requiring the knowledge of the material chemistry only. *Structural* feature parameters, such as the atomic coordination number and bonding distances, and radial distribution function, require knowledge of the specific atomistic structures of the material (in addition to the chemistry), and they need to be determined experimentally or from atomistic simulations, such as MD simulations in the present work.

For fast mapping the glass properties in a complex computational space, it is not efficient to use both the *chemical* and *structural* feature parameters to construct descriptors. Densities and elastic moduli of the $SiO_2$-based glasses are indeed strongly correlated with or determined by many of the glass structural features, such as atomic packing density, coordination numbers and ring sizes of network formers[5,7,53–58]. These *structural* feature parameters, however, are unknown for a given glass composition in the present work unless the MD simulations have been performed to obtain the corresponding atomistic structure. On the other hand, if the atomistic structure of a glass material is already known, there is no need to perform any ML-based predictions as the elastic moduli and density can be easily and quickly calculated via a molecular static simulation using the strain-stress method described in *Section 2*. In fact, obtaining the glassy structure via MD simulations is the most time-consuming step when computing the density and elastic moduli of a $SiO_2$-based glass. Thus, only the *chemical* quantities are considered for the construction of the descriptors for the ML model in the present work. As a result, the developed ML model is able to predict the properties by only requiring the information of the glass chemistry and without the need to run any additional MD simulations.

The construction of the descriptors should always start with the ones that are physically relevant to the material property of interest. Here, the parameters of the MD force-field potentials are chosen as the *chemical* feature parameters to generate the descriptors since the calculations of the atomic interactions of each MD run are based on these parameters according to Equation 1. The calculated density and elastic moduli are actually derived quantities from many multilevel and intricate MD runs. Therefore, these parameters could be a set of suitable candidates to construct the ML descriptors for the MD-calculated glass density and elastic moduli.

In the present work, the descriptors associated with the Coulomb interactions for a given glass composition is written as,

Equation 6
$$u^{Coul}_{q_m,q_n} = \sum_i c_{i_m} \cdot q_m \cdot \sum_j c_{j_n} \cdot q_n$$

where $q_m$ and $q_n$ denote the effective ionic charges listed in Table 2, which have values among -1.2, +0.6, +1.2, +1.8 and +2.4 *e*, and $c_{i_m}$ and $c_{j_n}$ denote the mole fractions of the constituent elements *i* and *j* with effective ionic charge $q_m$ and $q_n$, respectively. For example, for a glass that containing Na, K, Ca and Sr, as the effective ionic charges are +0.6 *e* for Na/K and +1.2 *e* for Ca/Sr, respectively (Table 2), the descriptor that corresponds to the Coulomb interactions between the ions with +0.6 and +1.2 *e* charges is calculated as $u^{Coul}_{+0.6,+1.2} = (c_{Na_{+0.6}} + c_{K_{+0.6}}) \cdot 0.6 \cdot (c_{Ca_{+1.2}} + c_{Sr_{+1.2}}) \cdot 1.2$, where $c_{Na_{+0.6}}$, $c_{K_{+0.6}}$, $c_{Ca_{+1.2}}$, and $c_{Sr_{+1.2}}$ are the elemental mole fractions of Na, K, Ca, and Sr, respectively. Because there are five different types of charge valences assigned for the elements that modeled in the present work, the total number of the Coulomb interactions descriptors, $u^{Coul}_{q_m,q_n}$, is $C_5^1 + C_5^2$=15.

As shown in Equation 1 and Table 2, the MD parameters associated with the Buckingham term describe the short-range interactions between each ion in a very complex way. Since we do not have *a priori* knowledge of how to combine these parameters to result in optimal modeling results, based on our previous experience[17], the corresponding descriptors are constructed as a series of weighted Hölder means, from which the ML model selects the most useful descriptors for modeling and predicting the glass properties of interest. As shown in Table 2, there are three individual Buckingham parameters (i.e., $A_{i,O}$, $B_{i,O}$ and $C_{i,O}$) for each element to describe its short-range interactions with the O anions (including the O-O self-interactions). Among these three parameters, the $B_{i,O}$ term influences the short-range interaction energy exponentially based on Equation 1. Therefore, different from $A_{i,O}$ and $C_{i,O}$, $B_{i,O}$ is not directly used as the feature parameter for the descriptor construction. Instead, in order to accurately describe the exponential effects of $B_{i,O}$, we proposed to use a parameter, $B'_{i,O}$, for the descriptor construction. The $B'_{i,O}$ parameter is calculated from $B_{i,O}$,

Equation 7
$$B'_{i,O} = \exp\left(-\frac{r^0_{i,O}}{B_{i,O}}\right)$$

where $r^0_{i,O}$ is the distance where the first derivative of the Buckingham form becomes zero. Therefore, for each type of the ions, $r^0_{i,O}$ is actually calculated from the values of $A_{i,O}$, $B_{i,O}$ and $C_{i,O}$. In addition, since $C_{i,O}$ of Li has a zero value, extra procedures were applied to obtain the value of the $r^0_{i,O}$ term for Li, which is described in detail in *Section 3 in Supplementary Material*. The calculated values of the $B'_{i,O}$ term for all the elements studied in the present work are summarized in Table 2, along with their MD parameters.

Thus, the descriptors associated with the short-range interactions are eventually generated from $A_{i,O}$, $B'_{i,O}$ and $C_{i,O}$ based on the glass composition ($c_i$) as the following,

Equation 8

$$u_p^x = \left(\sum_{i \in S_{ele}} c_i x_{i,O}^p\right)^{\frac{1}{p}}, p = -4, -3, -2, -1, 1, 2, 3, 4,$$

$$u_p^x = exp\left(\sum_{i \in S_{ele}} c_i \ln(x_{i,O})\right), p = 0,$$

where $u_p^x$ denotes the descriptors generated from the feature parameter $x$ associated with the Buckingham short-range interactions. There are three types of $x$, $A_{i,O}$, $B'_{i,O}$ and $C_{i,O}$. Let $S_{ele} = \{Si, O, Li, Na, K ...\}$ be the set of the elements contained in the glass. Different values of $p$ results in different Hölder means of $x$, which are the quartic-harmonic mean ($p = -4$), cubic-harmonic mean ($p = -3$), quadratic-harmonic mean ($p = -2$), harmonic mean ($p = -1$), geometric mean ($p = 0$), arithmetic mean ($p = 1$), Euclidean mean ($p = 2$), cubic mean ($p = 3$), and the quartic mean ($p = 4$), respectively. In addition, in Equation 8, $c_i$ is the mole fraction of the glass constituent element $i$, and $x_{i,O}$ is the value of the Buckingham parameter $x$ between the element $i$ and O. Besides, we also consider the standard deviation of the arithmetic means ($u_1^x$) as a type of descriptors, which is calculated as,

Equation 9

$$u_1^{x-\sigma} = \left(\left(\frac{1}{1 - \sum_{i \in S_{ele}} c_i^2}\right) \cdot \left(\sum_{i = i \in S_{ele}} c_i (x_i - u_1^x)^2\right)\right)^{\frac{1}{2}}$$

Based on Equation 8 and 9, thirty distinct descriptors are generated in total from $A_{i,O}$, $B'_{i,O}$ and $C_{i,O}$ (27 from Equation 8, and 3 from Equation 9). In addition, we include the multiplications between any two of the thirty descriptors as interaction terms to consider the non-linear relations among these descriptors. Finally, we also include the arithmetic mean of the atomic mass as an individual descriptor. As a result, overall 511 descriptors are generated for the ML models, in which there are fifteen descriptors associated with long-range Coulomb interactions, thirty descriptors generated from the MD parameters of the Buckingham term and 465 corresponding interaction terms (including self-interactions, thus $C_{30}^1 + C_{30}^2$=465), and one descriptor representing the mean atomic mass.

**3.2. Statistic models**

To leverage the training data as wisely as possible, two types of statistical learning models, namely the GBM-LASSO and the M5P regression tree model[31,32], were implemented in the present work to mathematically link the glass properties of interest (i.e., density, bulk and shear modulus) with the constructed descriptors. Young's modulus was not included as a learning target since it can be easily calculated from bulk and shear moduli based on Equation 5.

We first built the GBM-LASSO model using the gradient boosting machine (GBM) technique[30], which uses a gradient descent algorithm to iteratively produce a prediction model in the form of an ensemble of weak learning models. In the present work, the least absolute shrinkage and selection operator (LASSO)[29] method was employed to generate the weak learning model at each GBM iterative step. The LASSO method is able to select the important input descriptors by identifying the non-important descriptors with zero regression coefficients and meanwhile keep regresses regularly, especially when the simple linear regression model such as ordinary least square (OLS) does not work due to a relatively small sample size compared with the number of descriptors. As a result, the high-dimension problem (with many potential input descriptors) is simplified to a lower dimension or OLS problem. This method is particularly useful to address the regression problem in the present work, since the size of the training set is small so that the number of the input descriptors is almost the same as the number of the training data (~500). At each GBM iterative step, the LASSO method can both select the descriptors that are most relevant to the glass property being learned and perform an ordinal linear regression using the selected descriptors. In addition, a learning rate of 5% was used to attenuate the LASSO regression term at each GBM iterative step. Moreover, in order to avoid over-fitting the training data, our GBM-LASSO model was also implemented with a 10-fold cross-validation and a conservative risk criterion developed by de Jong et al.[17] to determine the optimal number of the GBM iterations.

As a comparison to the developed GBM-LASSO model, we also applied a widely used regression tree model, known as M5P and implemented in the Caret/Weka data mining packages[31,32], to the same training set. The M5P model was combined with a conventional decision tree model with the linear regression functions at the nodes. Specifically, the M5P model uses *all* of the descriptors for the linear regression performed at the tree nodes though it only uses partial descriptors for the tree establishment, which could be a problem when the number of the potential descriptors and the number of training data size are comparable. Therefore, in the present work, we first employ the M5P model to rank the importance of all the potential descriptors using the "varImp" function in the Caret package[32]. Then, the M5P model, including the final linear regression at each node, is run again with the top 100 descriptors that have been ranked from the first step. As a result, the number of descriptors used for the M5P model is comparable to the total number of the descriptors selected by the GBM-LASSO model. The tree structure of the present M5P model is optimized automatically using the prune function and 10-fold cross-validation resampling implemented in the Caret package[32]. For our specific learning problem, the M5P model has the advantage of being quickly trained.

## 4. Results and discussion
### 4.1. Regressions accuracy of training data

In the present work, the training dataset was generated by high-throughput MD simulations, which contains the densities, bulk and shear moduli (i.e., K and G) of 498 individual glass compositions in 11 binary and 20 ternary $SiO_2$-based systems as summarized in Table S5 in Supplementary Material. 11 types of additive oxides were considered, namely $Li_2O$, $Na_2O$, $K_2O$, CaO, SrO, $Al_2O_3$, $Y_2O_3$, $La_2O_3$, $Ce_2O_3$, $Eu_2O_3$ and $Er_2O_3$. The ML models were applied to learn each of the glass properties separately.

The densities from the MD-calculated training dataset are plotted in Figure 1 against the corresponding regression results from the ML models. For the sake of a clear representation, the data points are grouped into four categories, which are pure amorphous $SiO_2$, type-I glasses that only contain alkali and alkaline earth oxides as additives, type-II glasses that contain $Al_2O_3$ and other oxides, and type-III glasses that contain rare earth and other oxides. As shown in Figure 1, the glass densities produced from both GBM-LASSO and M5P models agree well with the results from MD calculations with root-mean-squared-errors (RMSE) as small as 0.0229 and 0.0325 g·cm$^{-3}$, respectively. It is also found that the distributions of the prediction residuals are close to norm distributions. Together with the histogram of residuals, Figure 1 implies the ML models demonstrate the correlations of interests very well without any abnormal performance. The regression results of the two ML models on the bulk and shear moduli are also illustrated as parity plots shown in Figures 2 and 3. Still, good agreements are observed between the predictions from ML models and those from MD simulations in the training set. The residuals of the models also approximately follow normal distributions. The regression RMSEs of K and G of the GBM-LASSO model are 2.99 and 1.31 GPa, respectively, while 2.59 and 0.97 GPa for the M5P model. In addition, the GBM-LASSO model seems to yield slight underestimations on the glass samples with higher moduli, as shown in Figures 2a and 3a.

Here, to further evaluate the regression accuracy of the ML models, we define the relative error as,

Equation 10    Relative error $= \frac{|X_{ML} - X_{MD}|}{X_{MD}}$    *(X=density, K or G)*

where $X_{MD}$ is the density or elastic modulus calculated from MD simulation and $X_{ML}$ is the prediction from the GBM-LASSO or M5P model. As shown in Table 3, for both K and G, over 60% of the predictions from both ML models have a relative error of less than 5%, and over 90% predictions are within a relative error of less than 10%, indicating that excellent regression accuracy is achieved. Additionally, we find that the LASSO method has indeed significantly shrunk the size of the descriptor set. Among the 511 input descriptors, only 119, 127 and 87 descriptors are found to have non-zero regression coefficients when the ML models predict the glass density, K and G, respectively. It is also found that many of these descriptors have been multiply used for the LASSO regressions at different GBM iterative steps, indicating they are indeed important and useful to describe these glass properties.

### 4.2. Prediction capability

Since the ML models are only trained with a small set of data from MD simulations for the binary and ternary systems, providing reliable predictions out of the domain of the training set is quite crucial for the present models in terms of the future applications in the practical glass design spaces. Here, we randomly choose 11 ternary, 30 quaternary, 30 quinary and 30 senary glass compositions that are not included in the training dataset to evaluate the prediction capabilities of the ML models in the compositional space beyond the training set. For each chosen composition, the GBM-LASSO and M5P models are applied to predict its density, K and G, and then MD simulations are correspondingly performed to validate the ML predictions. The validation results are shown as parity plots in Figure 4. In addition, the prediction errors are analyzed and summarized in Table 4 in the same way as the error analysis of the training process (Table 3). On the one hand, it is found that the M5P model seems to yield large uncertainties when extrapolating. As shown in Table 4, the RMSEs of the predictions from the M5P model with respect to MD validations are $0.1774 g \cdot cm^{-3}$, 5.24 and 2.27 GPa for density, K and G, respectively, which are much larger compared to the RMSEs of the learning results listed in Table 3 ($0.0325 g \cdot cm^{-3}$, 2.59 and 0.97 GPa for density, K and G). In addition, as shown in Figures 4a-4c, the data points in the parity plots of the extrapolative predictions are more scattered compared to the results of the training process (Figure 1c, Figure 2c and Figure 3c). Particularly, as marked out in Figure 4b and 4c, there are several predictions for the bulk and shear moduli that largely deviated from the MD results. Their relative errors are found to be over 20%. Moreover, it is worth to note that the M5P model is also trained by further decreasing the number of descriptors, which only resulted in a further increase in the training RMSEs but no significant improvements on the prediction RMSEs.

On the other hand, the developed GBM-LASSO model shows very promising prediction capabilities for the multicomponent glass systems beyond the training set. As shown in Figures 4d-4f, the density, K and G predicted from the GBM-LASSO model are in very good agreement with the MD results. Nearly 85% of the predictions for K and over 90% for G have relative errors less than 10%. Moreover, as shown in Table 4, the RMSEs of the predictions from the GBM-LASSO model with respect to MD validations are 0.0536 g·cm$^{-3}$, 3.69 and 1.34 GPa for density, K and G, respectively, agreeing well to the training uncertainties of the model listed in Table 3. The results suggest that, after training with a small set of data for only binary and ternary systems, the developed GBM-LASSO model shows promising abilities to give reliable predictions for multicomponent k-nary glasses as long as their constituent oxides are included in the training set.

Moreover, we find the prediction range of the GBM-LASSO model can be possibly expanded to cover more types of additive oxides by adding a small amount of related binary and ternary MD data to the training set. Here we use $B_2O_3$ and $ZrO_2$ as examples, as the Buckingham potentials for boron and Zr have been recently developed by Du et al.[34,45], which are also compatible with the set of MD potentials used in the present work. The original training set is slightly modified by adding a few new binary and ternary data with glass compositions containing $B_2O_3$ or $ZrO_2$. Specifically, 7 binary and 21 ternary data are added with compositions from the $xB_2O_3$-$(100-x)SiO_2$ (x = 5, 10, 15, 20, 25, 30 and 35) and $xB_2O_3$-$yNa_2O$-$(100-x-y)SiO_2$ systems (x,y=5, 10, 15, 20, 25 and 30, and x+y ≤ 35), respectively. Also, for $ZrO_2$, 13 new data are added to the training dataset, which are $xZrO_2$-$(100-x)SiO_2$ (x=5, 10, 15, 20, 25, 30, 35) and $xZrO_2$-$(35-x)Na_2O$-$65SiO_2$ (x=5, 10, 15, 20, 25, 30). The GBM-LASSO model is re-trained with the corresponding new training set. Notably, the density, K and G of the newly added glass compositions are well reproduced by the new training dataset, and the overall RMSEs are just slightly varied (0.012 g·cm$^{-3}$ for density, 0.26 GPa for K and 0.30 GPa for G) from the values listed in Table 3. As shown in Figures 5a and 5b, the non-linear effects of $B_2O_3$ on the bulk and shear moduli are accurately described for the $xB_2O_3$-$(100-x)SiO_2$ and $xB_2O_3$-$(30-x)Na_2O$-$70SiO_2$ glasses after training. Moreover, the newly trained model can then be expanded to predict for the multicomponent glasses that contain $B_2O_3$ and $ZrO_2$. As shown in Figure 5c, the ML predictions for several $B_2O_3$-containing compositions, which are not in the training set, are well confirmed by MD validations. Similar results are also observed for the $ZrO_2$-containing glasses as shown in Figure S4. These results suggest that the developed GBM-LASSO has great potentials to be further expanded to cover more types of additive oxides in the future. To achieve such expansions, we only need a few of MD simulations to generate the binary and ternary data containing new types of oxides for the training set.

We believe the outstanding prediction capability of the GBM-LASSO model may benefit from two aspects: the method of descriptor construction and the advantages of the regression algorithms employed in the model. As discussed in *Section 3.1*, instead of directly using the chemical composition as descriptors, the present model constructs descriptors from the compositional averages of the MD potential parameters. As a result, these descriptors can not only smoothly map the entire design space as they are continuous functions of the glass compositions but also contain the information to reflect the intrinsic physical features of each component element, which are compositionally discrete. More importantly, the construction method ensures that the total number of the descriptors is invariant to the arity of the glass chemistry. In other words, it generates the same amount of descriptors for any given glass composition, no matter how many types of additive oxides it contains, as long as the interatomic potentials based on Equation 1 is used for MD simulations. In addition, most of the descriptors still have non-zero values even when the investigated glass contains only one or two types of additive oxides. As a result, this would allow the ML models to transform the extrapolation problems in the chemical compositional space into interpolation-like problems in the constructed descriptor space based on both glass composition and MD force-field parameters.

Furthermore, the GBM-LASSO model may also benefit from some unique features of the regression algorithms employed in the model. In principle, a good prediction ability means a model should avoid over-fitting performance and still achieve a regression accuracy as high as possible. In the present work, due to a relatively small size of the train set, the number of descriptors is almost the same as the number of training data. This results in a potential risk of over-fitting if all the descriptors are considered equally strong and used for regression. The LASSO regression method could be particularly useful to resolve this issue as it screens out the nonsignificant descriptors by setting their coefficient to zero. As a result, the risk of over-fitting could be efficiently reduced as the regression is actually produced by a much smaller number of descriptors.

Moreover, for a broader comparison, we also applied our descriptors and training/testing data with other two typical ML models, a frequently used GBM regression tree model (GBM-RT) implemented in the XGboost package[59] and a model using the elastic net method[60] under the GBM framework (GBM-EN). The prediction performances of these two models are described in detail in *Section* 5 in Supplementary Material. Comparing the prediction performances of all the test ML models (i.e., GBM-LASSO, GBM-EN, GBM-RT and M5P), it is noticed that GBM-LASSO/EN models generally show better performance than the tree-based models when predicting beyond the training set. One possible reason could be that the GBM-LASSO/EN models conduct continuous regression functions (LASSO and EN) by considering all the observations/descriptors simultaneously at each GBM-iterative step, and they do not perform data classification like the tree-based model. As a result, the regression processes enforce more smoothness than the tree-based models in the functions mapping continuous descriptors to observations, especially when the size of the training set is small and the targeted responses are continuous functions of descriptors. On the other hand, tree-based methods usually require hard thresholds on the classification boundary. This requirement could result in large prediction uncertainties for the untrained sample if one or several input descriptors have values very close to the classification boundary, especially when the model itself is trained with a small set of data but used for extrapolative predictions. For this reason, the GBM-LASSO model proposed in the present work could be advantageous for many of materials problems.

In these cases, the properties of interests (e.g., density and elastic moduli) are reasonably continuous and smooth to the descriptors (e.g., compositions), but the training set is relatively small and established from the studies of sparse regions.

**4.3. Comparison between ML predictions and experimental measurements**

To further evaluate the model reliability, the predictions of the present GBM-LASSO model are validated with a large amount of experimental data across a multicomponent compositional space. Specifically, we collected the experimentally measured density and shear (G) and Young's (E) moduli from the Sciglass 7.12 database, which in turn were gathered from academic literature and patents published up to May 2014[61], for the $SiO_2$-based glasses containing the 12 additive oxides (i.e., $Li_2O$, $Na_2O$, $K_2O$, CaO, SrO, $Al_2O_3$, $Y_2O_3$, $La_2O_3$, $Ce_2O_3$, $Eu_2O_3$, $Er_2O_3$, and $ZrO_2$) that have been considered in the present work. When collecting the data, we constrained the composition of $SiO_2$ to be no less than 50 mol%. In comparison, it is worth to note all the glass compositions in our MD training dataset have no less than 65 mol% $SiO_2$. Overall 550 data points, including 142 binary, 303 ternary, 95 quaternary and 10 higher-order data (oxide components more than four), were collected for G; 1010 data points, including 231 binary, 464 ternary, 157 quaternary and 158 higher-order data, were collected for E; 4647 data points, including 1327 binary, 2483 ternary, 607 quaternary and 230 higher-order data, were collected for density. Moreover, about 30% of the data have the $SiO_2$ composition less than 65 mol%, which can serve as a validation to test the extrapolation capability of the present ML model in the compositional space. In addition, among these collected data, some of them can correspond to the same or very similar glass compositions, but they are gathered from different literature sources, as the density and elastic moduli for those compositions have been measured multiple times previously.

For each of the collected experimental data point, we took the corresponding glass composition to predict the G, E and density using our GBM-LASSO ML model and compare them with the experimental values. The predicted E is calculated from predicted K and G based on Equation 5. It is worth to mention that the GBM-LASSO model is still only trained with the MD training set, and the collected experimental data were not used for training. As shown in Figure 6, the validation results are characterized as 2D-hexbin plots with the ML predicted results versus the experimental values. It is found that the predictions from the GBM-LASSO model generally agree well with the experimental measurements. Compared to the experimental values, over 50% of the model predictions have relative errors less than 7%, and about 90% predictions are with relative errors less than 15% for both G and E. In terms of density, the predictions from the ML model yields even better agreement with experiments, where over 80% of predictions have relative errors less than 3% and 96% of predictions are with relative error less than 6%.

Besides the general agreement between the ML predictions and experimental data, as shown in Figure 6, it is noted that there are still scattered ML predictions that are largely deviated from the experimental values. After a careful analysis, we found that many of these prediction outliers should result from the inconsistency between the experimental data as they were gathered from different sources. In other words, the predictions of the ML model are in a good agreement with other sets of the experimental data with the glass compositions that are equal or close to the outliers. Here we show two typical examples as marked by the dashed-line circles in Figure 6a. One set of the data there corresponds to a measurement on the $Li_2O$-$SiO_2$ binary glasses with $Li_2O$ contents ranging from 26 mol% to 40 mol%, in which shear modulus of the glasses were reported to range from 5.71 to 13.79 GPa[62]. In contrast, at the corresponding compositions, the ML model predicted that the shear moduli should be about 31~33 GPa, which are actually in very good agreement with the results of experimental measurements on similar glass compositions from other two studies[63,64]. Another set of data marked by the circle in Figure 6 corresponds to a measurement on the $Al_2O_3$-$Y_2O_3$-$SiO_2$ glasses[65], where the ML model yields conflict predictions. However, in the meanwhile, the ML predictions on the $Al_2O_3$-$Y_2O_3$-$SiO_2$ glass systems are also confirmed by other experimental measurements[66–68] (More details are described in Table S4). In addition, we acknowledge that, for some of the prediction outliers in Figure 6, we still cannot have clear reasons as there are no other data available for comparison. These outliers can result from the inaccuracy of the MD simulations or the ML model when predicting the elastic moduli for some specific glass chemistries. For example, it is found that the present ML model generally underestimates the densities of ternary glasses containing both $Al_2O_3$ and rare-earth oxides (i.e., $Y_2O_3$, $La_2O_3$, $Eu_2O_3$ and $Er_2O_3$).

More importantly, after we remove these outliers (i.e. 15 out of 550, 35 out of 1010, and 77 out of 4647 in total for G, E and density, respectively) that can be confidently regarded as the experimental inconsistency, the RMSEs of the predictions from the present GBM-LASSO model are 2.51 GPa, 6.67 GPa and 0.0700 g·cm$^{-3}$ for G, E and D respectively, which are reasonably small by considering the possible uncertainties of the experimental measurements. Such uncertainties are quite common in the Sciglass database due to different experimental methods and sources (one example is shown in Figure S2b). The general agreements between the ML predictions and experimental data shown in Figure 6 further support the prediction reliability of the present GBM-LASSO model in a complex compositional space.

In addition, when validating with the experimental data for the $B_2O_3$-containing glasses from the Sciglass database, we found that the present GBM-LASSO model could have relatively large uncertainties in prediction accuracy. For example, the model predictions on the Young's moduli of the $B_2O_3$-$Na_2O$-$SiO_2$ ternary glasses are found to agree with the experimental measurements from some certain groups[69–71] (RMSE: ~6.33 GPa) but largely deviate from other experimental data in the Sciglass database (RMSE:~15.05 GPa)[61]. There are two possible reasons for such fluctuations in prediction accuracy. First, the experimental data from different studies already contain large fluctuations in elastic moduli for glasses with similar chemical compositions[72–74], indicating potentially large errors in some experiments. Second, the force-field potential of $B_2O_3$ employed in the present work can be inaccurate in terms of describing the elastic moduli. As reported by the developers of this $B_2O_3$ potential[45], the MD predicted bulk, shear and Young's modulus can be much higher than the experimental values in the $B_2O_3$-$Na_2O$-$SiO_2$ ternary system (up to 50% depending on the concentrations), although the variation trends with respect to the glass compositions are reproduced. However, because of the consistency between the MD results and our ML predictions (Figure 5), our developed GBM-LASSO model still has the capability to provide more reliable and accurate predictions for the $B_2O_3$-containing glasses, as long as compatible interatomic potentials that are more accurate on elastic properties are developed. Under that situation, one would only need to use the new interatomic potential to calculate a small amount of binary and ternary data and incorporate them into the training set.

Furthermore, the prediction capability of the GBM-LASSO model on elastic moduli is also evaluated by comparing it with a widely used physics-based model developed by Makishima and Mackenzie[75,76], hereafter referred to as MM model. Noteworthily, the MM model requires the actual density of the glass as an extra input, but the present GBM-LASSO model can make predictions only according to glass compositions, which makes it more suitable to be used as a fast screening tool before practical syntheses. Additionally, in the MM model, the interactions between atoms are assumed to be fully ionic so that Young's modulus can be derived from the Coulomb form of the electrostatic energy[76]. Such an ionic assumption could be problematic when it is applied for modeling the transition-metal oxides since the partially covalent characteristics of the metal-oxygen chemical bonds cannot be ignored. However, the covalent characteristics can be well captured by the Buckingham short-range interaction parameters in MD simulations, which are also used as input features to construct ML descriptors in the present work.

Indeed, compared to the MM model, it is found that the GBM-LASSO model yields considerable improvements on the elastic moduli predictions for the $SiO_2$-based glasses containing transition-metal oxides. By using an experimental validation dataset collected from the Sciglass database, which is composed of multicomponent $SiO_2$-based glasses with $Y_2O_3$ as one of the constituent components, the prediction RMSE of the GBM-LASSO model is calculated to be 10.16 GPa. As a comparison, the prediction RMSE of the MM model on the same dataset is as high as 22.42 GPa if the density-inputs are taken from the predictions of a widely used empirical regression model developed by Priven[10], and 13.39 GPa if experimental densities are used as inputs. Similar results were also observed for the $ZrO_2$-containing glasses, where the prediction RMSE of the GBM-LASSO model is 6.69 GPa, much smaller than that of the MM model, which is 10.55 GPa. More detailed information is shown in Figure S6 in Supplementary Material.

As a further demonstration, we also performed an investigation in the $Y_2O_3$-$SiO_2$ binary systems. Since there are no experimental measurements for this binary system, we performed *ab-initio* MD simulations (AIMD) on bulk modulus (K) for several glass compositions to validate the results of our classical MD simulations. Due to the high computational costs, the AIMD simulations were not performed for predicting Young's modulus. The calculation settings of the AIMD simulations are described in detail in *Section* 8 in Supplementary Material. As shown in Figure 7, the bulk modulus predicted from the GBM-LASSO model agree well with both the classical MD and AIMD simulations. However, the predictions from the MM model largely deviate from the results of MD simulations using the glass densities no matter computed from classical MD simulations or predicted from the widely used empirical model developed by Priven[10].

**4.4. Rapid screening of glass density and elastic moduli**

The GBM-LASSO model developed in the present work is able to predict the density and elastic moduli of a given glass composition in a negligible fraction of a second, making it possible for a rapid and comprehensive screening on these properties in a complex compositional space. As an illustration, we apply the trained GBM-LASSO model to systematically map the distributions and variations of the densities and elastic moduli of $Y_2O_3$-doped soda-lime-alumina glasses. Specifically, a quinary compositional space composed of $Na_2O$, $CaO$, $Al_2O_3$, $Y_2O_3$ and $SiO_2$ is homogenously meshed with a compositional interval of 1.0 mol% and under a constraint that the concentration of $SiO_2$ is no less than 65.0 mol%. The GBM-LASSO model is employed to predict the density, K and G for the glass composition at each mesh point. Overall, 82,251 compositions were studied by running the program on a regular personal computer (PC) in just a few hours. In contrast, tremendous computational powers ($10^8$ ~$10^9$ CPU hours) will be burned if purely using the MD simulations to generate the same amount of data.

The prediction results are visualized in Figure 8 as a 2D histogram plot with respect to density and Young's modulus, E, which is calculated from predicted K and G based on Equation 5. From a practical point of view, one would expect a structural glass to have Young's modulus as high as possible, and meanwhile keep a relatively low density. From Figure 8 we can know that most of the glasses in the $Na_2O$-$CaO$-$Al_2O_3$-$Y_2O_3$-$SiO_2$ system have Young's moduli around 83 GPa and densities around 2.6 g·cm$^{-3}$. From the screening, it is also found that low Young's moduli generally occur for the glasses with high $Na_2O$ contents, while the large additions of $Al_2O_3$ and $Y_2O_3$ result in a significant enhancement on Young's moduli, which is consistent with the previous experimental observation[66]. As marked by the red-dashed-line circle in Figure 8, one can achieve a series of glasses with Young's moduli higher than 100 GPa and densities ranging from 2.5 ~ 3.1 g·cm$^{-3}$ by optimizing the contents of the additive oxides. In addition, from the screening results, one can also know that it is probably difficult to prepare glasses with densities lower than 2.4 g·cm$^{-3}$ but Young's moduli larger than 80 GPa in this system. All in all, using the present developed GBM-LASSO model, a compositional-property database for any glass systems of interest can be rapidly generated as long as the corresponding force-field potentials are available and accurate enough to describe the structural and elastic properties. These databases allow the designers to have a fruitful overview on the density and elastic properties to enlighten their own design before experimental syntheses.

## 5. Conclusion

In this work, we demonstrated a novel machine-learning framework to efficiently learn and predict densities and elastic bulk and shear moduli of $SiO_2$-based glasses across a multicomponent compositional space, including 13 types of additive oxides, namely $Li_2O$, $Na_2O$, $K_2O$, CaO, SrO, $Al_2O_3$, $Y_2O_3$, $La_2O_3$, $Ce_2O_3$, $Eu_2O_3$, $Er_2O_3$, $B_2O_3$ and $ZrO_2$. Our framework combines a learning/predicting statistical model developed by implementing least absolute shrinkage and selection operator with a gradient boost machine (GBM-LASSO), high-throughput MD simulations to provide training data, and a diverse set of descriptors to generalize the chemistries of k-nary $SiO_2$-based glasses. Notably, the descriptors are constructed from the force-field potential parameters used for MD simulations so that they have the capability to bridge the empirical statistical modeling with the underlying physical mechanisms of interatomic bonding. Consequently, even training with a simple dataset only composed of binary and ternary glass samples, the developed GBM-LASSO exhibits promising prediction capability to allow for quick and accurate predictions on the density and elastic moduli for any k-nary glasses within the 14-component composition space. The prediction reliability of the developed GBM-LASSO ML model is evaluated by validating with a large amount (>>1000) of both simulation and experimental data. Furthermore, after comparing with other frequently used ML models, we found that the outstanding prediction capability of the GBM-LASSO model may benefit from both the way of descriptor construction and the advantages of the regression algorithms employed in the model. In addition, it is found that the GBM-LASSO model also yields considerable improvements on the elastic moduli predictions for the $SiO_2$-based glasses containing transition-metal/rare-earth oxides compared to the widely used MM model[75,76]. Such improvements originate from the capacity of our ML model to accurately describe the partially covalent bonding characteristics between the transition metal and oxygen atoms. Finally, as an example of its the potential applications, we utilized the model to perform a rapid screening on 82,251 compositions of a quinary glass system to construct a compositional-property database that allows for a fruitful overview on the glass density and elastic properties.

The present work is focused entirely on the modeling of glass density and elastic moduli; however, our ML framework could also be advantageous for the study of other glass physical properties and structural features. Our future studies will be a ML modeling on a few of fundamental glass structural properties, such as bridge/non-bridge oxygen ratio and angle distribution, ring size distributions of the network formers and average coordination number and bond length of cations, which are well-known to be essential to understand many of the physical and mechanical behaves of the $SiO_2$-based glasses. With the present work and more future works, a composition-structure-property database that sits nicely in the "Materials Genome Initiative" landscape[26,77–80] is desired to be developed via ML techniques and serve as powerful tools for the practical design of new glasses in the future. More generally, the methods of descriptor construction and the ML framework introduced in the present work could also be advantageous for many other materials science problems, where the datasets are of modest size and extrapolative predictions in high-dimensional space are required from the learning based on the low-dimensional sparse regions.

**Data availability**

The data of the MD training and test sets are summarized in Tables S5, S6 and S7 in Supplementary Material. The code that supports the findings of this study are available from the corresponding author (qiliang@umich.edu) on request.

**Competing interest statement**

The authors declare no competing financial or non-financial interests.

**Author Contributions Statements**

Y.J.H and L.Q. proposed the methodology of descriptors construction. G.Z. and Y.J.H. conceived and implemented the statistical machine learning models. Y.J.H. and M.Z. performed the high-throughput MD simulations. B.B. performed the training of the GBM-RT model. Y.J.H., T.D.R. and G.Z. performed the screening work. Q.Z., Q.Z., Y.C, X.S. and L.Q. conceptualized and supervised the research project. M.d.J. assisted and provided guidance on the computer programing of the machine learning models. Y.J.H, G.Z. and L.Q. prepared the manuscript. All authors discussed the results and contributed to the manuscript.

**Acknowledgement**

Y.J.H and Q.L. acknowledge support by the gift funding from Continental Technology LLC, Indianapolis, Indiana, USA. The high-throughput MD simulations were supported through computational resources and services provided by Advanced Research Computing at the University of Michigan, Ann Arbor. This work also used the Extreme Science and Engineering Discovery Environment (XSEDE)[81] Stampede2 at the TACC through allocation TG-DMR190035.

**Table 1** The $SiO_2$-based binary and ternary systems of which the density and elastic moduli are calculated by high-throughput MD simulations (marked in green). The calculated results are used as a training dataset for the ML models.

|  | $Li_2O$ | $Na_2O$ | $K_2O$ | CaO | SrO | $Al_2O_3$ | $Y_2O_3$ | $La_2O_3$ | $Ce_2O_3$ | $Eu_2O_3$ | $Er_2O_3$ |
|---|---|---|---|---|---|---|---|---|---|---|---|
| $Li_2O$ | ■ | ■ | ■ | ■ |  | ■ | ■ | ■ |  |  | ■ |
| $Na_2O$ |  | ■ | ■ | ■ | ■ | ■ | ■ | ■ | ■ | ■ | ■ |
| $K_2O$ |  |  | ■ | ■ |  |  |  |  |  |  |  |
| CaO |  |  |  | ■ | ■ | ■ |  |  |  |  |  |
| SrO |  |  |  |  | ■ |  |  |  |  |  |  |
| $Al_2O_3$ |  |  |  |  |  | ■ |  |  |  |  |  |
| $Y_2O_3$ |  |  |  |  |  |  | ■ |  |  |  |  |
| $La_2O_3$ |  |  |  |  |  |  |  | ■ |  |  |  |
| $Ce_2O_3$ |  |  |  |  |  |  |  |  | ■ |  |  |
| $Eu_2O_3$ |  |  |  |  |  |  |  |  |  | ■ | ■ |
| $Er_2O_3$ |  |  |  |  |  |  |  |  |  |  | ■ |

**Table 2** Effective ionic charge and Buckingham potential parameters used for MD simulations[25,33–42]. Here, $A_{i,O}$, $B_{i,O}$ and $C_{i,O}$ are the short-range interaction parameters between an ion element $i$ and oxygen anion. The short-range interactions between the cation elements are ignored in the present set of MD potentials. The values of $B'_{i,O}$, calculated based on Equation 7, is also listed for each element.

| Element | Effective ionic charge $q_i$ ($e$) | Buckingham potential parameters | | | |
|---|---|---|---|---|---|
| | | $A_{i,O}$ (eV) | $B_{i,O}$ (Å) | $C_{i,O}$ (eV·Å$^6$) | $B'_{i,O}$ |
| O[35,36] | -1.2 | 2029.22 | 0.343645 | 192.58 | 66.7013 |
| Si[35,36] | +2.4 | 13702.91 | 0.193817 | 54.681 | 105.6045 |
| Li[37] | +0.6 | 41051.94 | 0.15116 | 0 | 25.5680 |
| Na[35,36] | +0.6 | 4383.756 | 0.243838 | 30.7 | 34.0818 |
| K[37] | +0.6 | 20526.97 | 0.233708 | 51.489 | 17.8292 |
| Ca[25] | +1.2 | 7747.183 | 0.252623 | 93.109 | 49.3250 |
| Sr[25] | +1.2 | 14566.64 | 0.245015 | 81.773 | 26.8815 |
| Al[38] | +1.8 | 12201.42 | 0.195628 | 31.997 | 50.0620 |
| Y[39] | +1.8 | 29526.98 | 0.211377 | 50.477 | 20.9356 |
| La[38] | +1.8 | 4369.39 | 0.2786 | 60.28 | 30.2441 |
| Er[40] | +1.8 | 58934.85 | 0.195478 | 47.651 | 17.1005 |
| Eu[41] | +1.8 | 5950.529 | 0.253669 | 27.818 | 19.5874 |
| Ce[42] | +1.8 | 11476.95 | 0.242032 | 46.7604 | 21.8666 |
| B[45] | +1.8 | 12362.78* | 0.171271 | 28.500 | 164.7216* |
| Zr[34] | +2.4 | 17943.38 | 0.226627 | 127.65 | 58.3358 |

* $A_{i,O}$ and $B'_{i,O}$ values for the boron ions are calculated for the glass composition of 30% $B_2O_3$+70% $SiO_2$.

**Table 3** Regression results of the GBM-LASSO and M5P machine learning models on the training set, including root mean squared error (RMSE), and the percentage of predictions within 5, 10, 20, and 30 percent relative errors according to Equation 10, respectively. The units of RMSE are g/cm$^3$ and GPa for density and elastic moduli, respectively.

| Property | Model | RMSE | Percent of Predictions within Relative Error of | | | |
|---|---|---|---|---|---|---|
| | | | 2.5% | 5% | 10% | 20% |
| Density | GBM-LASSO | 0.0229 | 98.8 | 100.0 | 100.0 | 100.0 |
| | M5P | 0.0325 | 96.6 | 100.0 | 100.0 | 100.0 |
| K | GBM-LASSO | 2.99 | 33.9 | 61.8 | 91.0 | 99.6 |
| | M5P | 2.59 | 40.6 | 70.1 | 94.6 | 99.6 |
| G | GBM-LASSO | 1.31 | 47.4 | 76.3 | 96.0 | 100.0 |
| | M5P | 0.97 | 57.6 | 89.8 | 99.4 | 100.0 |

**Table 4** Prediction errors of the GBM-LASSO and M5P machine learning models for the glass compositions that are not included in the training set, including root mean squared error (RMSE), and the percentage of predictions within 5, 10, 20, and 30 percent relative error according to Equation 10, respectively. The tested compositions are from 11 ternary, 30 quaternary, 30 quinary and 30 senary systems that are randomly chosen. The units of RMSE are g/cm$^3$ and GPa for density and elastic moduli, respectively.

| Property | Model | RMSE | Percent of Predictions within Relative Error of | | | |
|---|---|---|---|---|---|---|
| | | | 2.5% | 5% | 10% | 20% |
| Density | GBM-LASSO | 0.0536 | 86.1 | 99.0 | 100.0 | 100.0 |
| | M5P | 0.1774 | 62.4 | 80.2 | 93.1 | 97.0 |
| K | GBM-LASSO | 3.69 | 34.7 | 51.5 | 83.2 | 100.0 |
| | M5P | 5.24 | 28.7 | 48.5 | 78.2 | 96.0 |
| G | GBM-LASSO | 1.34 | 41.6 | 76.2 | 98.0 | 100.0 |
| | M5P | 2.27 | 36.6 | 57.4 | 90.1 | 97.0 |

**Figure 1** Performances of the ML models on the glass densities of the training set. (a) Performance of the GBM-LASSO model. (b) Distribution of residuals between the GBM-LASSO predictions and the MD results of the training set. (c) Performance of the M5P model. (d) Distribution of residuals between the M5P predictions and the MD results of the training set. The curved lines in (b) and (d) are normal distributions constructed from the mean and standard deviation of the residuals. The data points are grouped into four categories based on their glass chemistry, which are pure amorphous $SiO_2$, type-I glasses that only contain alkane and alkane earth oxides as additives, type-II glasses that contain $Al_2O_3$ and other oxides, and type-III glasses that contain rare earth and other oxides.

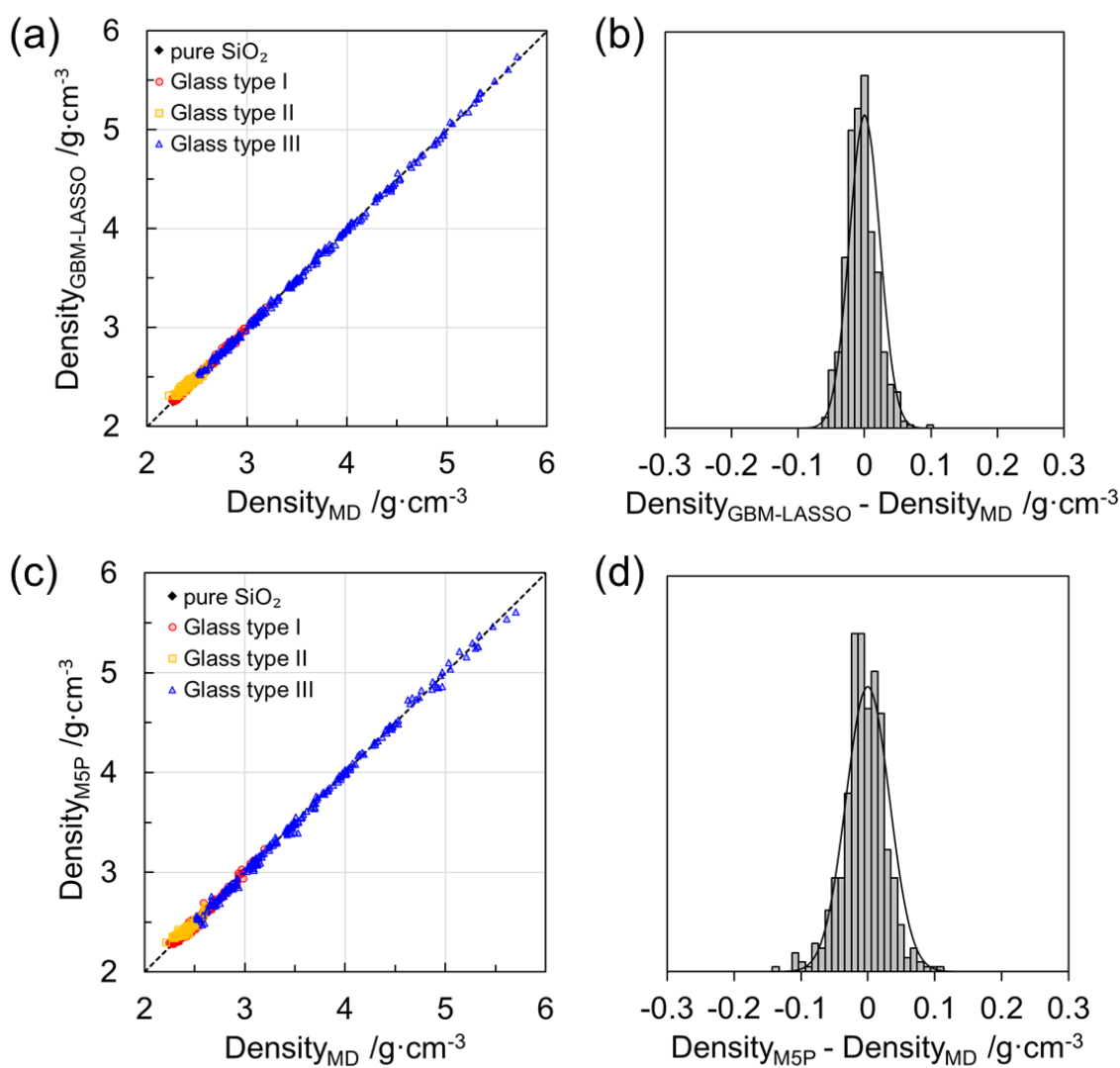

**Figure 2** Performances of the ML models on the bulk moduli (K) of the training set. (a) Performance of the GBM-LASSO model. (b) Distribution of residuals between the GBM-LASSO predictions and the MD results of the training set. (c) Performance of the M5P model. (d) Distribution of residuals between the M5P predictions and the MD results of the training set. The curved lines in (b) and (d) are normal distributions constructed from the mean and standard deviation of the residuals. The data points are grouped into four categories based on their glass chemistry by following the definitions Figure 1.

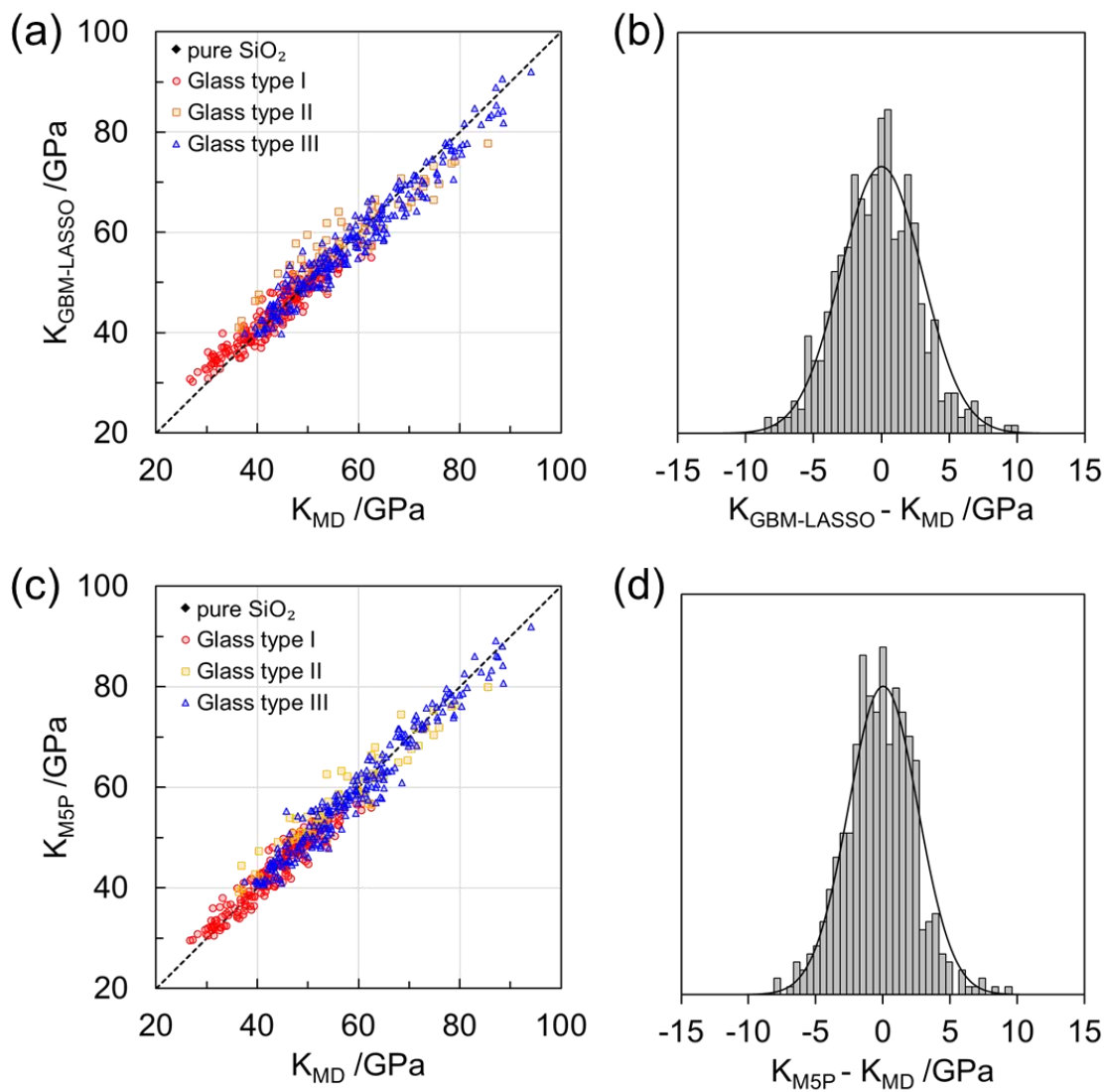

**Figure 3** Performances of the ML models on the shear moduli (G) of the training set. (a) Performance of the GBM-LASSO model. (b) Distribution of residuals between the GBM-LASSO predictions and the MD results of the training set. (c) Performance of the M5P model. (d) Distribution of residuals between the M5P predictions and the MD results of the training set. The curved lines in (b) and (d) are normal distributions constructed from the mean and standard deviation of the residuals. The data points are grouped into four categories based on their glass chemistry by following the definitions in Figure 1.

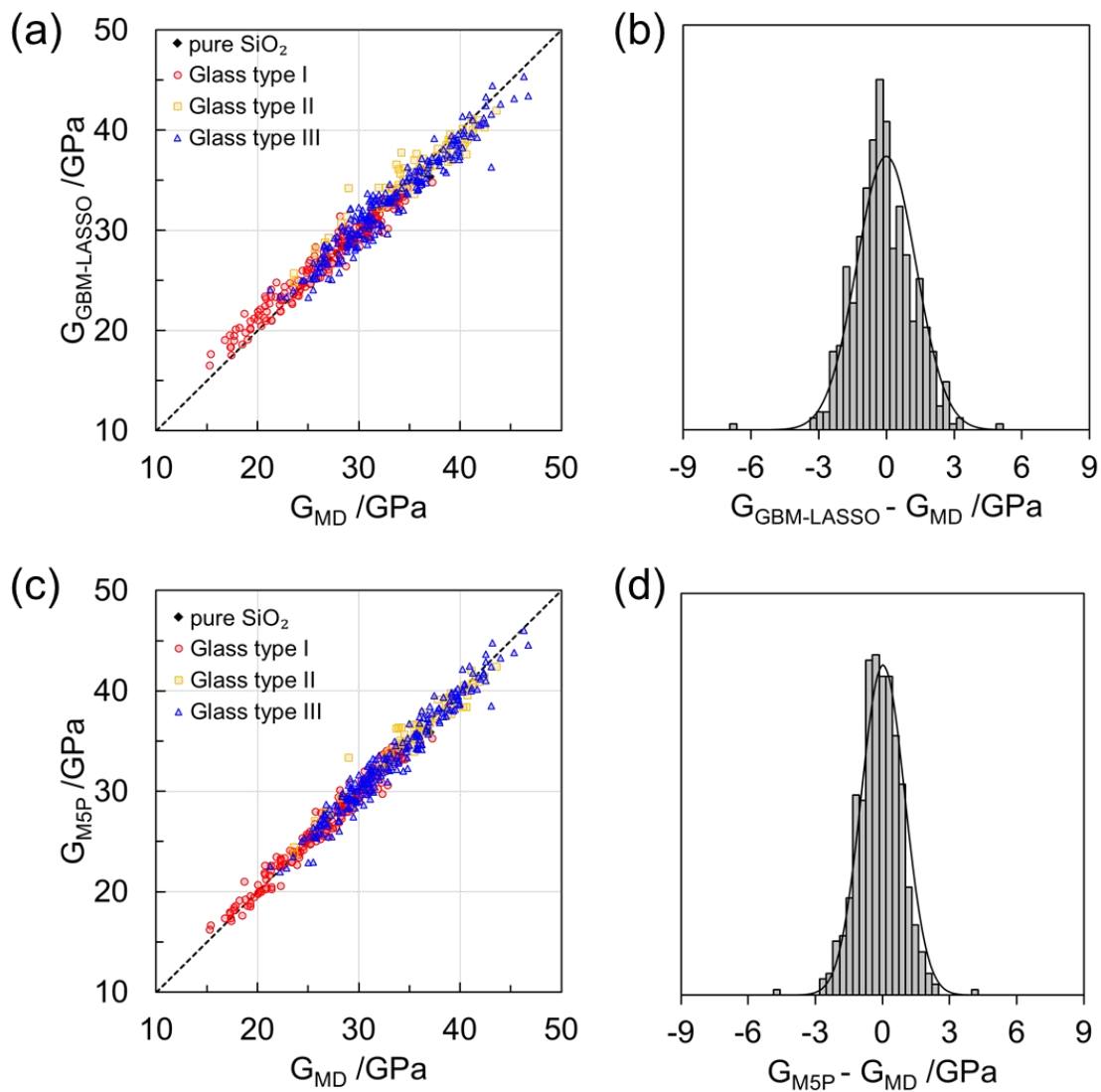

**Figure 4** Prediction performance of the ML models on the glass compositions beyond the training set. The predictions from the M5P and GBM-LASSO model are plotted *versus* the validation results from MD simulations. (a)-(c): Density, bulk and shear moduli predicted by the M5P model. (d)-(f): Density, bulk and shear moduli predicted by the GBM-LASSO model. The glass compositions used for the testing are from 101 randomly chosen ternary, quaternary and senary systems that are not included in the training set. The composition information of each data point can be found in Supplementary Material (Table S6). The data points within the black dot-dashed region have relative errors less than 10%.

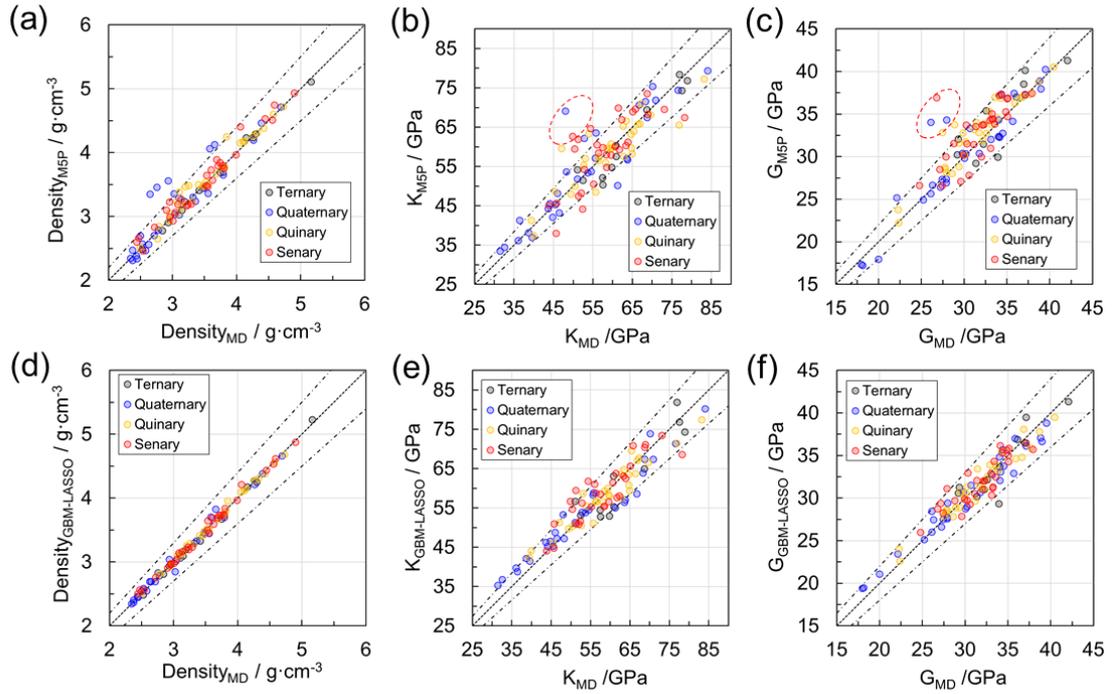

**Figure 5** Prediction performance of the GBM-LASSO model after adding data of new oxide species (e.g., $B_3O_2$) to the training set. (a)-(b) Reproduction of the non-linear effects of $B_2O_3$ on bulk and shear modulus in the (a) $xB_2O_3\text{-}(100\text{-}x)SiO_2$ and (b) $xB_2O_3\text{-}(30\text{-}x)Na_2O\text{-}70SiO_2$ glasses in the training set. (c) Predictions from GBM-LASSO *versus* MD results on the test set. The test set is composed of 15 randomly selected compositions for the $B_2O_3$-containing multicomponent glasses (detailed information is listed in Table S7) that are not included in the training dataset.

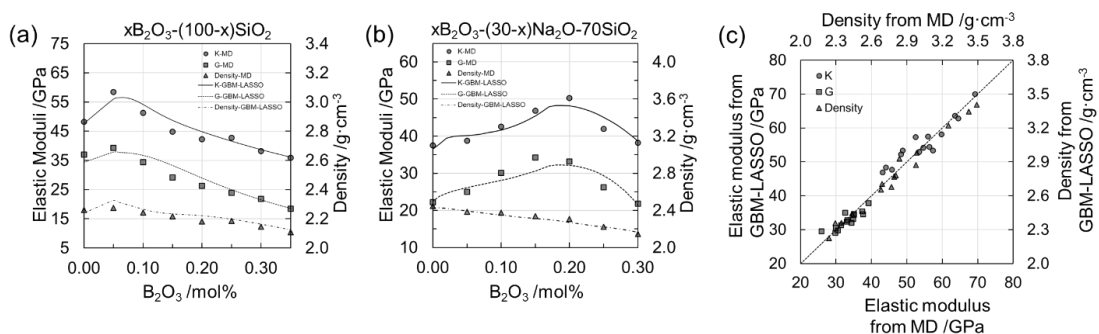

**Figure 6** Glass properties predicted by the GBM-LASSO model *versus* the experimental values reported from the Sciglass 7.12 database[61]. The GBM-LASSO model is only trained with the MD training set and the experimental data were not used for training. (a) Shear modulus; (b) Young's modulus; (c) Density. The dashed line is the identity where the predictions are equal to the experimental values. The hexagonal unit with a hotter color means that there are more data points within the coverage area of the unit. The dashed-line circles in Figure 6a mark out typical examples of prediction outliers caused by experimental data inconsistency.

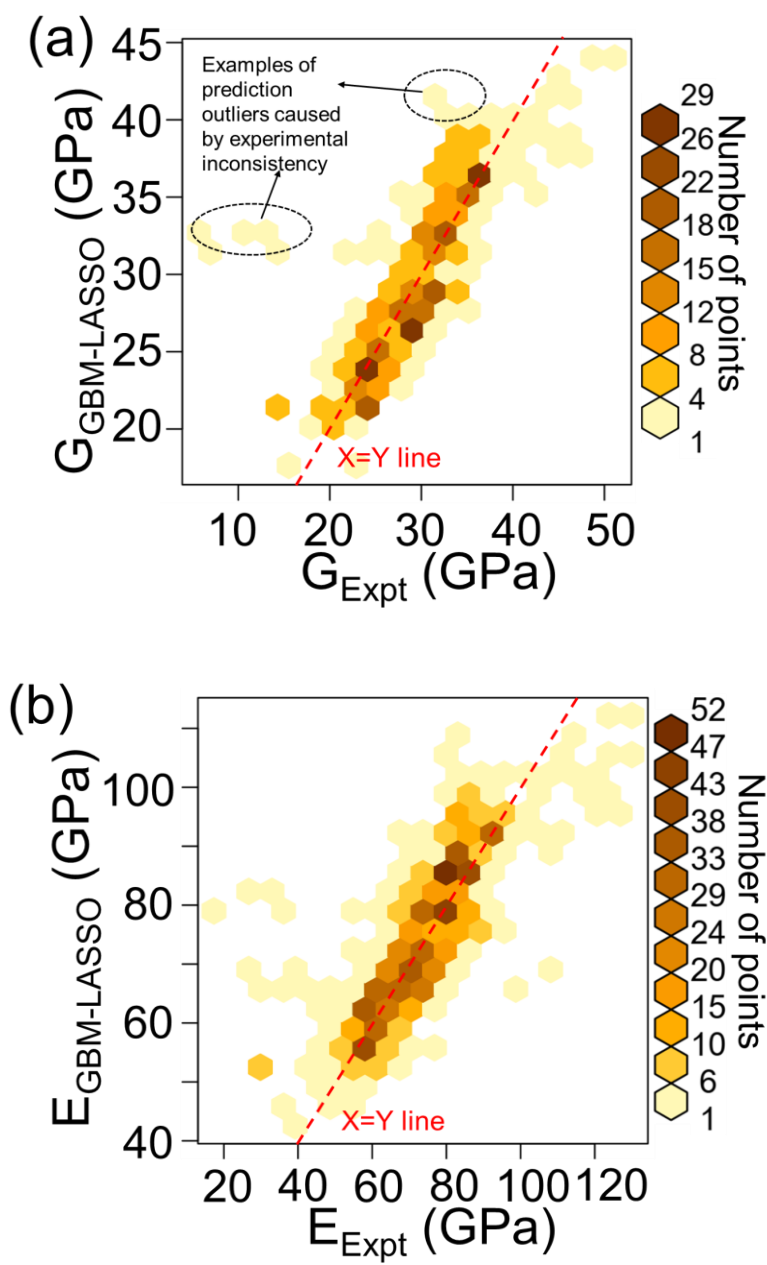

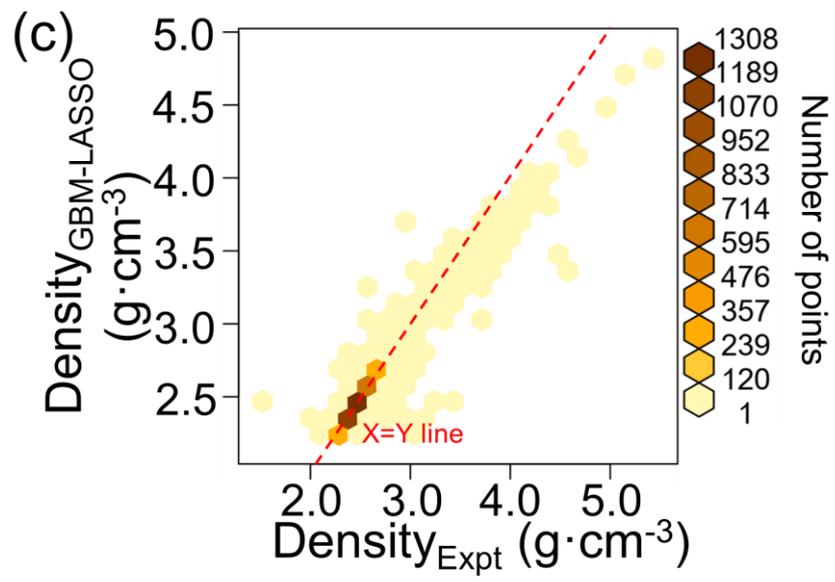

**Figure 7** Bulk modulus of the $Y_2O_3$-$SiO_2$ binary glasses calculated from the classical MD and AIMD simulations and predicted from the GBM-LASSO and MM models[75]. The error bar of the AIMD results are generated from the results calculated under different applied strains.

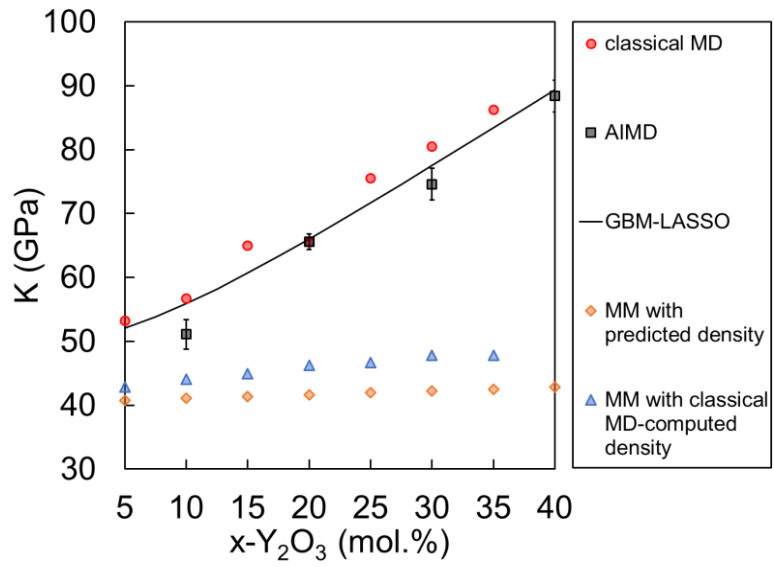

**Figure 8** A 2D histogram to visualize the distributions of the density and Young's modulus (E) of the glasses in the $Na_2O$-$CaO$-$Al_2O_3$-$Y_2O_3$-$SiO_2$ system. The histogram is generated from 82,251 compositions, where the GMB-LASSO model is employed to predict the density and Young's modulus. The content of $SiO_2$ is constrained to be no less than 65 mol%. The hexagonal unit with a hotter color means that there are more glass compositions having density and E within the coverage area of the unit.

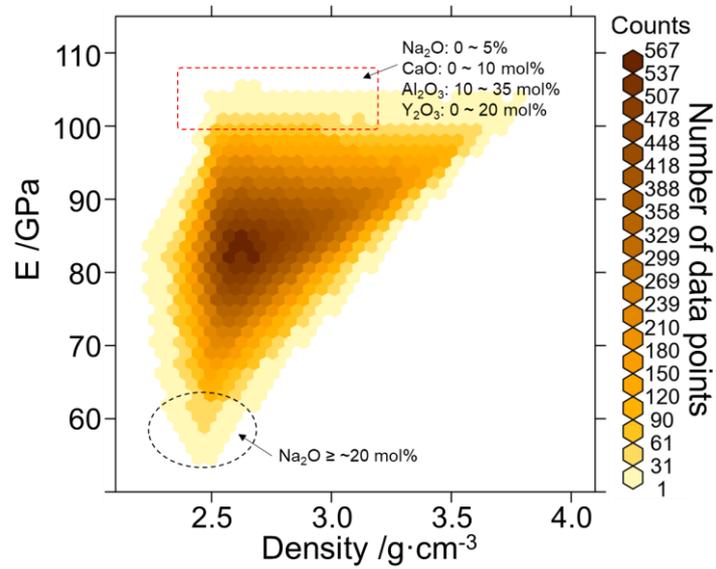